\documentstyle[preprint,aps,epsfig]{revtex}
\newcommand{\vecr}{{\bf r}}

\newcommand{\bm}{\bibitem}
\newcommand{\ud}{\bf}

\begin{document}

\draft
\tighten

\title{Coulomb and nuclear breakup of $^{8}$B}
\date{\today}
\author{R. Shyam}
\address {Saha Institute of Nuclear Physics, Calcutta - 700 064, India\\
E-mail address: shyam@tnp.saha.ernet.in}
\author{I.J. Thompson}
\address{Department of Physics, University of Surrey, Guildford GU2 5XH, U.K.\\
E-mail address: I.Thompson@surrey.ac.uk}
\maketitle

\begin{abstract}
The cross sections for the ($^8$B,$^7$Be-$p$) breakup reaction on
$^{58}$Ni and $^{208}$Pb targets at the beam energies
of 25.8 MeV and 415 MeV have been calculated within a one-step
prior-form distorted-wave Born approximation. 
The relative contributions of Coulomb and nuclear breakup of dipole
and quadrupole multipolarities as well as their interference have been
determined. The nuclear breakup contributions are found to be
substantial in the angular distributions of the $^7$Be fragment  
for angles in the range of 30$^\circ$ - 80$^\circ$ at 25.8 MeV beam
energy. The Coulomb-nuclear interference terms make the dipole
cross section larger than that of quadrupole even at this low beam energy.     
However, at the incident energy of 415 MeV, these effects are almost
negligible in the angular distributions of the ($^7$Be-p)
coincidence cross sections at angles below 4$^\circ$.

\end{abstract}

\section{Introduction}
The Coulomb dissociation (CD) method provides an alternate
indirect way to determine the cross sections for the radiative
capture reaction $^7$Be($p,\gamma$)$^8$B at low relative
energies~\cite{baur94,baur97}, which is the most important
and most uncertain nuclear input to the standard solar model
calculations~\cite{bahc89,turc93}. The CD method reverses
the radiative capture by the dissociation of a projectile (the
fused system) in the Coulomb field of a target by making the
assumption that nuclei do not interact strongly
and the electromagnetic excitation process is dominated by 
a single multipolarity.

Motobayashi et al.\  performed the first measurement of the
breakup of $^8$B into $^7$Be-$p$ low energy continuum
in the field of $^{208}$Pb with a radioactive $^8$B beam of 
46.5 MeV/A energy~\cite{moto1}, which was analyzed by
us~\cite{shyam96} using Alder-Winther's semiclassical theory
of Coulomb excitation (which assumes the colliding nuclei are 
point-like objects)~\cite{alder}. Considering only the $E1$
component of the excitation, we found~\cite{shyam96} the
measured breakup cross sections
to be consistent with a $S_{17}$(0) = (15.5 $\pm$ 2.80)
eV barn.  However, in the CD of $^8$B at these energies, the $E2$
component of breakup may not be negligible~\cite{lang94},
and its presence affects the extracted  $S_{17}$(0) \cite{shyam96}.
Although some authors~\cite{gai95} dispute this claim, a
reliable estimate of this component is necessary for the accurate
extraction of the astrophysical S-factor ($S_{17}$(0)) from 
the breakup data.

Since the contribution of the $E2$ component is strongly dependent
on the nuclear structure model of $^8$B (which is not yet known with
certainty), the RIKEN group has repeated the breakup measurements
with angular distributions extended to larger scattering angles
where the cross sections are more sensitive to the $E2$
component~\cite{moto2}. An analysis of this data within the
distorted wave Born-approximation (where the breakup is treated as
the inelastic excitation of the projectile to the continuum) led
these authors to conclude that the $E2$ components and the nuclear
breakup effects are considerably smaller. However, they  
use a collective model prescription to calculate the inelastic
nuclear form factor (see eg.~\cite{shyam96}).  Due
to a long tail in the $^8$B g.s wave function this procedure is
unlikely to be accurate. Furthermore, Coulomb breakup is calculated
by a point-like projectile approximation (PLPA) in these studies, and its
range of validity is yet to be determined for this projectile.

The Notre Dame group has measured the breakup of $^8$B on the
$^{58}$Ni target at the beam energy of 25.8 MeV, well below the
Coulomb barrier, where the $E2$ component is expected to dominate
the CD process~\cite{nd}. However, the reliable
extraction of the $E2$ component from this data, where only the
the integrated cross section of the $^7$Be fragment is measured,
is still doubtful. The analysis of the data reported in Ref.~\cite{nd}
used the Alder-Winter's semiclassical theory of Coulomb excitation,
where the final state is treated as a two-body
system, thus assuming that the measured angles of $^7$Be were
equal to those of the $^7$Be-$p$ center of mass. The inadequacy
of this assumption has been demonstrated in~\cite{shyam97}.
Furthermore, the total breakup cross section reported in this
experiment could not be reproduced  within the Alder-Winther
theory even if a wide variety of structure models of $^8$B were
used~\cite{nunes1}. On the other hand, the
importance of the nuclear breakup effects in the kinematical
regime of this experiment has recently been
emphasized~\cite{nunes2}. Therefore, there is a need to reanalyze
this data using a proper theory where 
the three-body kinematics is taken into account. 
 
In this paper we perform a one-step prior-form DWBA analysis of the
$^8$B breakup data at both low and high energies in order to check
the validity of various assumptions of the Coulomb dissociation
method. We describe the breakup process as a single proton excitation
of the projectile from its ground state to a range of states in the
continuum, which is discretized by the method of continuum bins. 
 
In previous calculations~\cite{nunes1,nunes2} based on this
model, quadrupole transitions from the ground state to the continuum
did not include the excitation of $f$ partial waves in all cases. Moreover, the 
numerical calculations were performed only for the breakup reaction 
$^8$B + $^{58}$Ni $\rightarrow$ $^7$Be + $X$ at the
beam energy of 25.8 MeV where the final state was considered as 
as a two-body system. In this work we improve upon this aspect by
using a proper three-body kinematics for the final state, and also include
excitations of $\ell$ = 3 continuum states everywhere.

We furthermore apply our model to the RIKEN data where $^7$Be and $p$ were
detected in coincidence with very small relative energies in a
$^8$B induced reaction on $^{208}$Pb target at the beam energy of 
415 MeV. We avoid the point-like projectile approximation as well
as collective model prescription for the nuclear form factor,
by determining the nuclear and Coulomb parts by a single-folding 
method where the relevant fragment-target interactions are folded
by the projectile wave functions in the ground and continuum states.

We indicate our formalism in the next sections. Details of the numerical
calculations are described in section III. The results of our
calculations and their discussions are presented in section IV, while 
the summary and conclusions of our work are given in section V.  
 
\section{Formalism}
\subsection{DWBA reaction mechanism}
The cross section for the breakup of a projectile
($a$) into its fragments ($c$ and $p$), $t + a \rightarrow c + p + t$,
can be represented to first order 
by the inelastic excitation of the projectile $a$ to
its continuum, $t + a \rightarrow a^{*} + t$ by means of the 
 `prior form' DWBA matrix element. This is an integral of the form 
\begin{eqnarray}
T_{DWBA}^{(-)} & = & <\phi_{{\bf k}_{cp}}(\vecr)  \chi_{a^*t}^{(-)}({\bf p}_{a^*},{\bf R})
                     \mid V_{ct}(\vecr_c)+V_{pt}(\vecr_p) \mid
                     \phi_0({\bf r})
                     \chi_{at}^{(+)}({\bf p}_a,{\bf R})>.
\end{eqnarray}
In this expression $\phi_0$ is the ground state wave function of
the projectile, $\chi_{at}^{(+)}({\bf p}_a,{\bf R})$
describes the relative motion of the projectile with 
respect to the target, and 
$\phi_{k_{cp}}(\vecr) \chi_{a^*t}^{(-)}({\bf p}_{a^*},{\bf R})$
describes the excited continuum state ${\bf k}_{cp}$
of the projectile fragments after
breakup. The superscripts $(+)$ and $(-)$ specify the 
outgoing or incoming wave boundary conditions satisfied by the 
corresponding wave functions respectively.
The coordinates ${\bf R}$ and ${\bf r}$ describing the motion of
the CM of the fragments with respect to the target and their 
relative motion respectively are related to those of the relative 
motion of the individual fragments from the target
(${\bf r}_p$ and ${\bf r}_c$) by  
\begin{eqnarray} \label{coords}
{\bf r}_p & = & {\bf R} + \frac{m_c}{m_a}{\bf r}, \nonumber\\
{\bf r}_c & = & {\bf R} - \frac{m_p}{m_a}{\bf r}.
\end{eqnarray}
The $V_{ct}(\vecr_c)$ and $V_{pt}(\vecr_p)$ are the core ($^7$Be) and
proton optical potentials, respectively, for interaction with the target.

Alternatively, the DWBA may be regarded as the first iteration of a
perturbative solution of a coupled-channels problem, and the breakup
cross sections found from the S-matrix elements of the outgoing
channels when a set of coupled equations is iterated once. The double
differential cross section for a $c-p$ breakup state from ground state
spin $j_a$ into final spin state $j_{a^*}$ and relative momentum
$k_{cp}$ is given by
\begin{eqnarray} 
\frac{d^2\sigma_\ell(k_{cp})}{dk_{cp}d\Omega_{a^*}} & = &
   {1\over {(2j_a + 1)}}
   \sum_ {m'  m}
            |\sum_ {LL'J}
    \langle L0 ~ j_a m | Jm\rangle  
    \langle L'm-m' ~ j_{a^*} m' | J m\rangle  \nonumber \\
&\times&     {{4 \pi}\over p_a} ~ \sqrt {{{p_{a^*}\over p_a}}}
    ~ e ^ {i( \sigma _ L + \sigma' _ {L'} )}    ~ {i\over 2}
     S^{J}_{Lj_a, L'j_{a^*} k_{cp}}
%    ~ \sqrt {2L+1\over {4 \pi}}~ Y_{L'} ^{m-m'}(\theta , \phi )  | ^2 ,
    ~ Y_{L} ^{0}(0,0)~ Y_{L'} ^{m-m'}(\Omega_{a^*})  | ^2 
,
\end{eqnarray}
where $p_a$ is the incident momentum of $a$ in the $\bf z$ direction
and $\Omega_{a^*}$ is the direction of momentum ${\bf p_{a^*}}$ of the
CM of the $c-p$ system with respect to the target nucleus $t$. In Eq.
(3), $J$ denotes the total angular momentum of the system and $L$,
$L'$ are the initial and final angular momentum of the CM of $c-p$ pair
with respect to the target.  In this work the S-matrix method has been
used, and the following paper \cite{nunes99} continues this approach to
present solutions of the full coupled equations, though only at the
lower (Notre Dame) sub-Coulomb energy.

Since our approach is based on a first-order solution of the coupled-channels
equations, we refer to the accompanying paper \cite{nunes99} (section II),
for details of these equations and of the coupling
matrix elements. The DWBA results presented below arise from the
solution of Eq.(5) of \cite{nunes99} for $n=1$. This implies that
the entrance and exit channel potentials for the 
$\chi_{at}^{(+)}({\bf K}_i,{\bf R})$ and $\chi_{a^*t}^{(-)}({\bf K}_f,{\bf R})$
channel wave functions are not optical potentials, but are
the diagonal interactions found by  
folding the fragment potentials $V_{ct}(\vecr_c)+V_{pt}(\vecr_p)$
over the initial or final states. These states are either
the initial state $\phi_0$ or the continuum bin states (\cite{nunes99}, Eq. 6)
for the bin centers and widths to be described below.

\subsection{Coincidence cross sections}

The {\em triple} differential cross section for the breakup of a projectile
$a$ into its fragments $c$ and $p$ can be related to the {\em double}
differential cross sections shown in \cite{nunes99} (for the
inelastic excitation of the projectile $a$ to its continuum state ${\bf k}_{cp}$). This is accomplished, as in \cite{shyam97}, by 
\begin{eqnarray} \label{tbk}
\frac{d^3\sigma}{dE_{c}d\Omega_{c}d\Omega_{p}} & = & \frac{J^\prime}{4\pi}
                       \frac{d^2\sigma}{dE_{cp}d\Omega_{a^*}}
                         \frac{\partial E_p}{\partial E_{tot}},
\end{eqnarray}
where the total kinetic energy
\begin{eqnarray}  \label{energy}
E_{tot} & = & E_c + E_p + E_t  \nonumber \\
        & = & E_{cp} + E_{a^{*}} + \frac{P^2}{2(m_a + m_t)}.
\end{eqnarray}
is related to the projectile energy ($E_a$) and the reaction
Q-value ($Q$) by
$E_{tot} \, = \, E_{a} + Q$. In Eq. (\ref{energy}) $E_c$, $E_p$,
and $E_t$ are the
kinetic energies of the fragments $c$, $p$ and recoiling target nucleus
respectively, while $E_{cp}$ and $E_{a^{*}}$ are the kinetic energies
of the relative motion of the fragments and of their CM with respect
to the target nucleus respectively. In Eq. (\ref{tbk}), we have assumed
that the angular distribution of fragments is isotropic in the projectile
rest frame; the expressions without making this assumption are given in
Ref.~\cite{Bau89}. The last factor in Eq. (\ref{tbk}) is given by
\begin{eqnarray} \label{ederiv}
\frac{\partial E_p}{\partial E_{tot}} & = & m_t \left [m_p + m_t - m_p
                     \frac{{\bf p}_p \cdot ({\bf P} - {\bf p}_c)}
                     {p_{p}^{2}} \right ]^{-1},
\end{eqnarray}
%		     - {\mbox{\boldmath $ p $}}_{c})}
and the Jacobian $J^\prime$ is defined as
\begin{eqnarray} \label{jacob}
J^\prime & = & \frac{m_c p_c m_p p_p}{\mu_{cp} p_{cp} \mu_{at} p_{a^{*}}} \ .
\end{eqnarray}
In Eqs. (\ref{coords}, \ref{ederiv}, \ref{jacob}), $m_{i}$ is the mass of the fragment $i$, and
$\mu_{at}$ and $\mu_{cp}$ are the reduced masses of the $a-t$ and
$c-p$ systems respectively. $\bf{P}$ is the total momentum which
is fixed by the conditions in the entrance channel.
The momenta ${\bf p}_{cp}$ and ${\bf p}_{a^{*}}$,
describing the relative motion of the fragments $c$ and $p$ and the
motion of their CM with respect to the target nucleus respectively,
can be related to their individual momenta ${\bf p}_c$ and ${\bf p}_p$
by straight-forward expressions (see. e.g. Ref.~\cite{Fuc82,shyam97}).
% The cross section $d\sigma/dE_{cp}d\Omega_{a^*}$, shown in \cite{nunes99},
% describes the inelastic excitation of the
% projectile from its ground state to the state ${\bf k}_{cp}$
% in the continuum.

\section{The numerical procedure and application to $^8$B breakup}

The $^8$B nucleus has a $2^+$ ground state at --0.137 MeV,
which is composed predominantly of the $^7$Be core in its
$3/2^-$ ground state and a valence proton in a $p_{3/2}$ 
configuration. $E1$ ($\lambda$=1) and $E2$ ($\lambda$=2) 
mechanisms will populate the ($s_{1/2}$, $d_{3/2}$, $d_{5/2}$) and
($p_{1/2}$, $p_{3/2}$, $f_{5/2}$, $f_{7/2}$) single particle
continuum states respectively in the first step.  We consider
the excitation to states of these partial waves up to
$E_{p-^7Be}$ = 3 MeV.

We have taken a single particle model for the structure of $^8$B
assuming that all states in $^8$B are determined by the g.s. potential
defined in~\cite{esb}. This simplification of the $^7$Be-$p$ scattering
state interaction, neglecting the core couplings and the M1 transitions, 
has hardly any effect on the integrated CD cross section (~\cite{nunes1})
at non-relativistic energies.

As remarked earlier, in the DWBA calculations we have to use 
diagonal channel potentials for the entrance and exit channels.
Different choices for these potentials are 
(a) pure $^8$B+target Coulomb, (b) Coulomb + some fixed optical
potential, or (c) using the monopole ($\lambda=0$) parts of the single-folded
potentials (eq. (15) of \cite{nunes99}). 
Choice (a) is appropriate for comparison 
with simple (e.g. semiclassical) models. Choice (b) is 
inaccurate for breakup states, as can be seen by comparison
with the very diffuse potentials found in method (c).  Choice (c)
has the advantage that it generalizes readily when all the
diagonal/off-diagonal
monopole/multipole couplings need to be calculated for a full CDCC
calculation as in \cite{nunes99}.

For the  Notre Dame experiment on the Coulomb dissociation of $^8$B on
$^{58}$Ni at 26 MeV~\cite{nd}, good accuracy for the continuum
discretization is obtained if we use 13 bins per partial wave, defined
in the following way: $9$ bins of $100$ keV centered at $0.15; 0.25; ...;
0.95$ MeV and $4$ bins of $500$ keV centered at 1.25; 1.75; 2.25; 2.75 MeV. 
Sufficient convergence for the $E1$ and $E2$ transitions was obtained if the
maximum number of the partial waves ($l_{max}$) and the maximum radius
($R_{max}$) were taken to be $600 \hbar$ and $300$ fm respectively.
Coupled asymptotic wave functions~\cite{CRCWFN}
are used beyond 30 fm.

For the  RIKEN experiment on the Coulomb dissociation of $^8$B on
$^{208}$Pb at 415 MeV~\cite{moto2}, 
good accuracy for the continuum discretization
is obtained if we use 8 bins per partial wave, defined in the
following way: $4$ bins of
$250$ keV centered at 0.125; 0.375; 0.625; 0.875 MeV and $4$ bins of $500$
keV centered at 1.25; 1.75; 2.25; 2.75 MeV. 
The bin wave functions were each integrated with 200 $k$-steps, 
out to $R_m=50$ fm.  In order to obtain convergence for the $E1$
and $E2$ transitions in this case, we include up to $l_{max}=10000
\hbar$ and $R_{max}=1000$ fm for the reaction mechanism.
Coupled asymptotic wave functions~\cite{CRCWFN} are used beyond 50 fm.

\section{Results and discussion}

In Figs. 1a and 1b, we show the angular distributions of $^7$Be and
$^8$B$^*$ respectively in a $^8$B induced breakup reaction on $^{58}$Ni
target at the beam energy of 25.8 MeV. Pure Coulomb and pure nuclear 
breakup cross sections are represented by the dashed and dashed-dotted 
curves respectively. The cross sections obtained by summing
coherently the Coulomb and nuclear amplitudes (to be referred as {\it total}
in the following) are represented by
the solid lines. In these calculations the procedure of single-folding
the respective fragment-target interactions with $^8$B ground and
continuum state wave functions (Eq. (11)) have been used. Nuclear parts
of $p$-$^{58}$Ni and $^7$Be-$^{58}$Ni interactions have been taken
from Refs.~\cite{becc} and~\cite{glov} respectively. The curves in 
Fig. 1b improve on those in \cite{nunes2} by the inclusion of $f$-wave
final states.

The angular distributions of $^7$Be and
$^8$B$^*$ are distinctly different from each other. While pure Coulomb and
{\it total} breakup cross sections show a forward peak in case of $^7$Be
(which is typical of the angular distribution of fragments emitted in
breakup reactions), those of $^8$B$^*$ tend to zero as angle goes to zero.
The latter is the manifestation of the adiabatic cut-off typical of the
Coulomb-excitation process. 
In both the cases the nuclear effects are small below 20$^\circ$ and there
is a Coulomb-nuclear interference minimum between 25$^\circ$ - 60$^\circ$.
However the magnitude of various cross sections are smaller in Fig. 1a.
Furthermore, the nuclear-dominated peak occurs at different angles in
Figs 1a ($\simeq$ 55$^\circ$) and 1b ($\simeq$ 70$^\circ$). As discussed 
in section (2.1), the angles of $^7$Be can be related to those
of $^8$B$^*$. A given $\theta_{7_{Be}}$ gets contributions from a range
of generally larger $\theta_{8_{B^*}}$. This explains to some extent the
shifting of the peaks of various curves to lower angles in Fig. 1a
as compared to the corresponding ones in Fig. 1b. This underlines
the important of three-body kinematics in describing the inclusive
breakup reactions.

The ratio of the experimental integrated breakup cross section of $^7$Be
(obtained by integrating the breakup yields in the angular range,
(45 $\pm$ 6)$^\circ$, of the experimental setup) to Rutherford elastic
scattering of $^8$B is reported to be
(8.1 $\pm$ 0.8$^{+2.0}_{-0.5}$) $\times$ $10^{-3}$~\cite{nd}. It is
not possible to get this cross section by
directly integrating the angular distributions shown in Fig. 1b 
in this angular range as the corresponding angles belong to $^8$B$^*$
and not to $^7$Be. However, in the three-body case (Fig. 1a), this
can be done in a straight-forward way. This gives a value of 
7.0 $\times$ 10$^{-3}$ which is in close agreement with the
experimental data. Thus, previous failures to explain the 
experimental value may be attributed to the neglect of both the
Coulomb-nuclear interference effects and the three-body
kinematics.

In Fig. 2, we have investigated the range of the validity of the point-like
projectile approximation (PLPA) and the role of the Coulomb-nuclear 
interference effects on the cross sections of dipole and quadrupole components
for the reaction discussed in Fig. 1a. In Fig. 2a the results for  
pure Coulomb breakup are shown. Dipole and quadrupole components of the cross
section obtained by the single-folding procedure are shown by solid and 
dashed lines respectively, while those obtained with the PLPA 
by solid and dashed lines with solid circles. It can be noted that
PLPA is not valid for angles beyond 20$^\circ$. The condition that
the impact parameter of the collision is larger than the sum of the 
projectile and target radii ($b > R_a + R_t$), 
assumed in applying the Alder-Winther theory, is no longer valid because 
there is a long tail in the $^8$B ground state wave function. We also 
note that the quadrupole component is affected more by the PLPA as compared
to the dipole. The big difference in the dipole and quadrupole cross sections
seen in the PLPA results beyond 20$^\circ$ (where the quadrupole component is
much bigger than the dipole), almost disappears in the
corresponding cross sections obtained by single-folding procedure. 
Nevertheless, the quadrupole cross sections still remain larger than those
of the dipole beyond 30$^\circ$ in the latter case.

In connection with PLPA, it should be made clear that $p$ + target and
the $^7$Be + target potentials {\em do} take into account the finite size of
the $^7$Be and target nuclei. This effect, however, is only important
when two nuclei are very close to each other and is masked by the 
nuclear effects which would be important at those impact parameters.

Dipole and quadrupole cross sections for pure nuclear breakup are shown
in Fig. 2b. The cross sections obtained by summing coherently
the amplitudes of $E1$ and $E2$ components of pure Coulomb and pure 
nuclear breakup are shown in Fig. 2c. We shall still refer the
corresponding components of the ${\it total}$ cross sections as 
E1 and E2, although usually the dipole and quadrupole
components of the ${\it pure}$ Coulomb cross sections are referred as 
such.  We notice that the Coulomb-nuclear interference effects 
make the contributions of the dipole component of the ${\it total}$ 
cross section larger than those of quadrupole one at all the angles.
This result is quite remarkable as it implies
that the $E2$ component of the total break up cross section in the
$^8$B induced reaction on $^{58}$Ni target is not dominant even
at the subCoulomb beam energies.  Therefore, there is hardly any hope
of determining the $E2$ component of $^8$B breakup by Notre Dame type
of experiment~\cite{nd}.

This underlines the need for more refined experiments to 
determine the $E2$ component (as already pointed out in
Ref.~\cite{lang94}). It is clear from Fig. 2c that the  
measurements of the angular distributions may provide  
useful information about the $E2$ component as it is  
different from that of the $E1$ multipolarity.
On the other hand, the angular
distributions of the fragments, calculated within a semiclassical
theory without making the approximation of isotropic angular
distributions in the projectile rest frame, have been shown
to have large $E1$ - $E2$ interference effects~\cite{esbe95,esbe96}.  
They lead to asymmetries in the momentum distributions of 
the fragments, whose measurements may enable one to put
constraints on the $E2$ component~\cite{davids}. However,
for the better accuracy of this method, improved
calculations including the nuclear effects may be necessary.
 
Our results for the nuclear effects in the angular
distribution of $^8$B$^*$ are approximately similar to
that reported in~\cite{vitturi}, where Coulomb and nuclear form
factors are calculated by folding the 
proton-target mean-field (parameterized by a Woods-Saxon function)
by the ground and discretized continuum state $^8$B wave functions.
These authors calculate various cross sections by integrating 
a fixed projectile-target optical potential along a semiclassical
trajectory. However, since the three-body kinematics for the
final state has not been considered by them, a direct 
comparison between their calculations and the data of~\cite{nd}
is not possible. 

In Fig. 3a, we show $E1$ and $E2$ components of the angular distributions
for the $^8$B + $^{208}$Pb $\rightarrow$ $^8$B$^*$ + $^{208}$Pb
reaction measured by the Kikuchi et al.~\cite{moto2} at the beam
energy of 415 MeV, for the pure Coulomb excitation case.
The dashed, dotted and solid lines represent $E1$, $E2$ and $E1+E2$
cross sections respectively which are obtained by the single-folding
procedure. Also shown in this figure are the corresponding results 
obtained by PLPA (curves with solid circles). We note that PLPA 
becomes inaccurate beyond 4$^\circ$ in this case.
Moreover, the $E2$ component of the pure Coulomb excitation becomes
increasingly important also after this angle.

In Fig. 3b, the {\it total} cross sections  
are shown. The nuclear part of the fragment-target interaction 
at these energies have been taken from~\cite{becc} (for proton)
and~\cite{zell} (for $^7$Be).
The dashed and dotted lines show the dipole and quadrupole cross
sections respectively, while the solid line represents their sum. 
It can be noted that nuclear effects modify the pure Coulomb
$E1$ cross sections substantially after $\sim$ 4$^\circ$, and  
the $E2$ cross sections in the entire angular range.
However, since the $E2$ components are quite small at angles
$\leq$ 4$^\circ$, the difference between pure Coulomb and
{\it total} dipole + quadrupole cross sections is appreciable only after
this angle.

Therefore, at RIKEN energies, the PLPA breaks down beyond 4$^\circ$,
where the Coulomb-nuclear interference effects as well as the quadrupole 
component of breakup is substantial. Hence, the Coulomb dissociation
method as used in e.g.  Ref.~\cite{shyam96} to extract a reliable
$S_{17}(0)$ from the measurements of the angular distributions in
the breakup of $^8$B on heavy target at RIKEN energies
($\sim$ 50 MeV/nucleon), is useful only when data is taken at
angles below 4$^\circ$. 

In Figs. 4a, 4b and 4c we show the comparison of our calculations
for $\epsilon \cdot d\sigma/d\theta$ with the experimental data of
Kikuchi et al.~\cite{moto2} as a function of the scattering angle
$\theta_{8_{B^*}}$ of the excited $^8$B (center of mass of the $^7$Be+$p$
system) for three relative energy bins. We have used the efficiency
($\epsilon$) matrix as well as angular and energy averaging as 
discussed in Ref.~\cite{moto2} which is provided to us by the
RIKEN collaboration. The dashed and dotted lines are the pure Coulomb
$E1$+$E2$ and $E2$ cross sections respectively while the solid
and dashed lines are
the corresponding {\it total} cross sections. We note that our
calculations are in fair agreement with the experimental data. We
stress that no arbitrary normalization constant has been used in 
the results reported in this figure.  

The quadrupole component of breakup is significant at almost all the  
angles in the relative energy bin 2.0 -- 2.25 MeV(c),
and at angles beyond  5$^\circ$ in the energy bin 1.25 -- 1.50 MeV(b).
On the other hand, its contribution is inconsequential in the energy
bin 0.5 -- 0.75 MeV (a). This result is in somewhat disagreement 
with that reported in Ref.~\cite{moto2}, where this component
is reported to be small everywhere below 1.75 MeV relative energy.
Although these authors also perform a quantum mechanical calculation
within DWBA, their treatment of the continuum state is very different
from ours. Moreover they use a collective model prescription for
the Coulomb and nuclear form factors, which has a limited
applicability for $^8$B breakup. Bertulani and Gai~\cite{bert98}
have also reported smaller quadrupole component in their analysis of this
data. These authors do not include the nuclear effects in the $E1$
excitations and make use of the eikonal approximation to
calculate the quadrupole nuclear excitation amplitudes. Moreover, the
Coulomb excitation amplitudes have been calculated with the 
PLPA which we have found to be invalid at higher angles (see Fig. 3).
It is also noted in Fig.3 that Coulomb-nuclear interference
effects reduce the $E1$ cross sections at larger angles.

Some authors have studied the importance of the higher order
effects in the Coulomb breakup of $^8$B~\cite{typel94,esbe95a,esbe96,typel97}. 
At RIKEN energies these effects play only a minor role for
this reaction in the kinematical regime of forward angles and low
relative energies~\cite{typel94,esbe96,typel97}. Therefore, our 
conclusions about the RIKEN data are unlikely to be affected 
much by the higher order breakup effects. However, the multi-step
breakup could play an important role at Notre Dame energies, which
is discussed in the following article~\cite{nunes99}.  

Finally we would like to remark that since we have adopted the 
single particle model of Ref.~\cite{esbe96} for the structure of $^8$B 
in our dynamical nuclear reaction calculations, the astrophysical
S-factor for the radiative capture reaction $^7$Be(p,$\gamma$)$^8$B
resulting from our analysis of the RIKEN data is the same as that of
these authors (17 eV.barn). It is should be noted that unlike the 
pure Coulomb dissociation calculations this is not a fitting parameter
in our analysis.

\section{Summary and Conclusions}

In this paper, we studied the breakup reactions $^8$B + $^{58}$Ni
$\rightarrow$ $^7$Be + X and $^8$B + $^{208}$Pb $\rightarrow$ $^8$B$^*$
+ $^{208}$Pb, at beam energies of 25.8 and 415 MeV respectively, 
within the framework of the one-step prior-form distorted wave
Born-approximation. In this theory, the breakup process is described
as a single proton excitation of the projectile from its ground state
to a range of states in the continuum, which is discretized by the
method of continuum bins. In this method, both Coulomb and nuclear breakup as well
as their interference terms are calculated within the same framework. 
Moreover, we use the three-body kinematics while
calculating the cross sections
for the $^7$Be fragment in the first reaction.

For the breakup reaction at low energy the Coulomb-nuclear interference
effects are found to be quite important. This leads to a reduction 
in the pure Coulomb dissociation cross sections for the 
$^7$Be fragment in the angular range of the measurements reported
in~\cite{nd}, which together with the three-body kinematics 
reproduces the experimental integrated cross section for this fragment.
This agreement had eluded the pure Coulomb dissociation calculations. 

A very striking feature of the Coulomb-nuclear interference
effect is that it makes the E1 component of the ${\it total}$ cross
section of the breakup reaction $^8$B + $^{58}$Ni $\rightarrow$ $^7$Be + X
(at the beam energy of 25.8 MeV), larger than the corresponding E2 component 
at all the angles. This renders untenable the main objective of
the Notre Dame experiment of determining the
$E2$ component in the breakup of $^8$B at low beam energies.
The dominance of the $E2$ component for this reaction at this energy,
seen in the semi-classical
Alder-Winther theory of Coulomb excitation has led to this expectation.
However, we note that even in pure Coulomb dissociation process, with 
finite size of the projectile taken into account, the $E2$
components is almost equal to that of $E1$ in the relevant angular range.

The breakup data at higher beam energies (RIKEN energies), are 
almost free from the nuclear effects and are dominated by the $E1$
component for $^7$Be-$p$ relative energies $<$ 0.75 MeV at very
forward angles ($\leq$ 4$^\circ$). The study of the breakup of
$^8$B in this kinematical regime is, therefore, better suited for the
extraction of reliable $S_{17}(0)$ for the capture reaction
$^7$Be($p,\gamma$)$^8$B at low relative energies.

\acknowledgements
One of the authors (RS) would like to acknowledge an associateship 
award from the Abdus Salam International Centre of Theoretical 
Physics, Trieste. This work is supported by Engineering and Physical 
Sciences Research Council, UK, grant nos. J/95867 and L/94574.

\newpage
\begin{center} {\bf Figure Captions} \end{center}
\begin{itemize}
\item[Fig. 1] 
Angular distribution of the $^7$Be fragment emitted in the
breakup reaction of $^8$B on $^{58}$Ni target at the beam energy of 25.8
MeV. The dashed and dashed-dotted lines show the pure Coulomb and
pure nuclear breakup cross sections respectively while their
coherent some is represented by the solid line. (b) Angular distribution
of $^8$B$^*$ in the Coulomb excitation of $^8$B on $^{58}$Ni at
the beam energy of 25.8 MeV. The dashed and dashed-dotted
lines show the cross sections for pure Coulomb and pure nuclear excitation
respectively, while the solid line represents their coherent sum. 

\item[Fig. 2]
Dipole (solid lines) and quadrupole (dashed lines) components of the
angular distributions of the $^7$Be 
fragment emitted in the breakup reaction of $^8$B on $^{58}$Ni target at
the beam energy of 25.8 MeV. (a) pure Coulomb breakup; also
shown here are the $E1$ (solid lines with solid circles) and $E2$
(dashed lines
with solid circles) cross sections obtained with point-like projectile
and target approximation (Alder-Winther theory), (b) pure
nuclear breakup and (c) Coulomb plus nuclear breakup where the
corresponding amplitudes are coherently summed.

\item[Fig. 3]
Angular distribution for $^8$B+$^{208}$Pb $\rightarrow$ $^8$B$^*$($^7$Be+p)+
$^{208}$Pb reaction at the beam energy of 415 MeV.
(a) Results for pure Coulomb excitation, the dashed and
dotted curves represent the $E1$ and $E2$ cross sections while their sum is
depicted by the solid line. Also shown here are the results obtained with a
point-like projectile and target approximation (Alder-Winther theory), where
dashed and dotted lines with solid circles show the corresponding
$E1$ and $E2$
cross sections while the solid line with solid circles represents their
sum. (b) Coherent sum of Coulomb and Nuclear excitation calculations;
the dashed and dotted lines show the dipole and quadrupole components while
the solid line is their sum. 

\item[Fig. 4]
Comparison of experimental and theoretical cross section
$\epsilon d\sigma/d\theta$ as a function of
the scattering angle $\theta_{8_{B^*}}$ 
for $^8$B+$^{208}$Pb $\rightarrow$ $^8$B$^*$($^7$Be+p)+
$^{208}$Pb reaction at the beam energy of 415 MeV.
Results for three relative energy bins of (a) 500-750 keV,
(b) 1250-1500 keV, (c) 2000-2250 keV are shown. $\epsilon$ is 
the detector efficiency. Solid lines show the calculated total Coulomb plus
nuclear dissociation cross sections while the dashed lines represents the
corresponding pure Coulomb dissociation result. Pure quadrupole  Coulomb and
Coulomb+nuclear cross sections are shown by dotted and dashed-dotted lines.
The experimental data and the detector efficiencies are taken
from~\protect\cite{moto2}. 
\end{itemize}
 
\begin{figure}
\begin{center}
\mbox{\epsfig{file=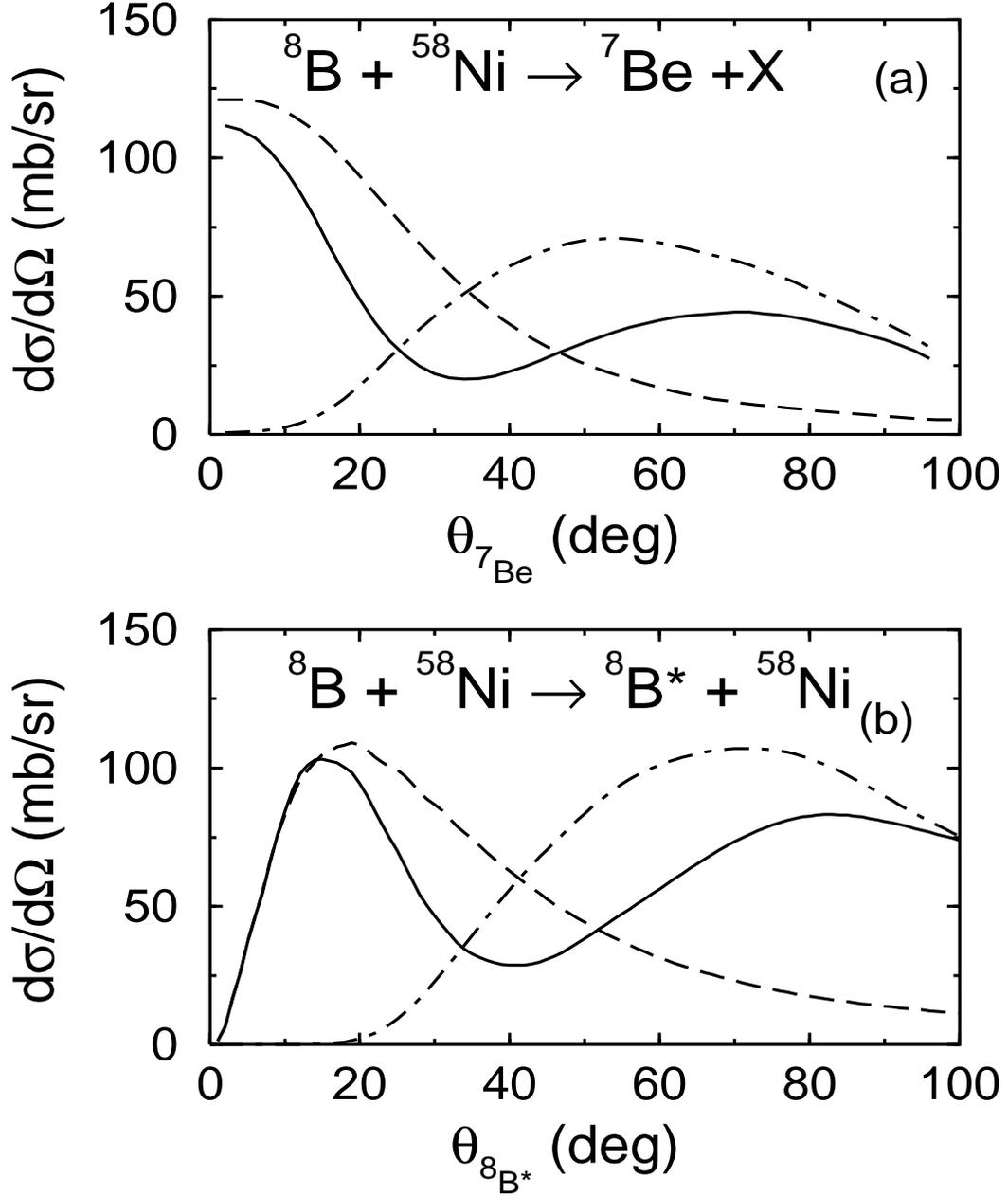,height=17.0cm,width=14.0cm}}
\end{center}
\caption{ (a) Angular distribution of the $^7$Be fragment emitted in the
breakup reaction of $^8$B on $^{58}$Ni target at the beam energy of 25.8
MeV. The dashed and dashed-dotted lines show the pure Coulomb and
pure nuclear breakup cross sections respectively while their
coherent some is represented by the solid line. (b) Angular distribution
of $^8$B$^*$ in the Coulomb excitation of $^8$B on $^{58}$Ni at
the beam energy of 25.8 MeV. The dashed and dashed-dotted
lines show the cross sections for pure Coulomb and pure nuclear excitation
respectively, while the solid line represents their coherent sum. } 
\label{fig:figa}
\end{figure}

\begin{figure}
\begin{center}
\mbox{\epsfig{file=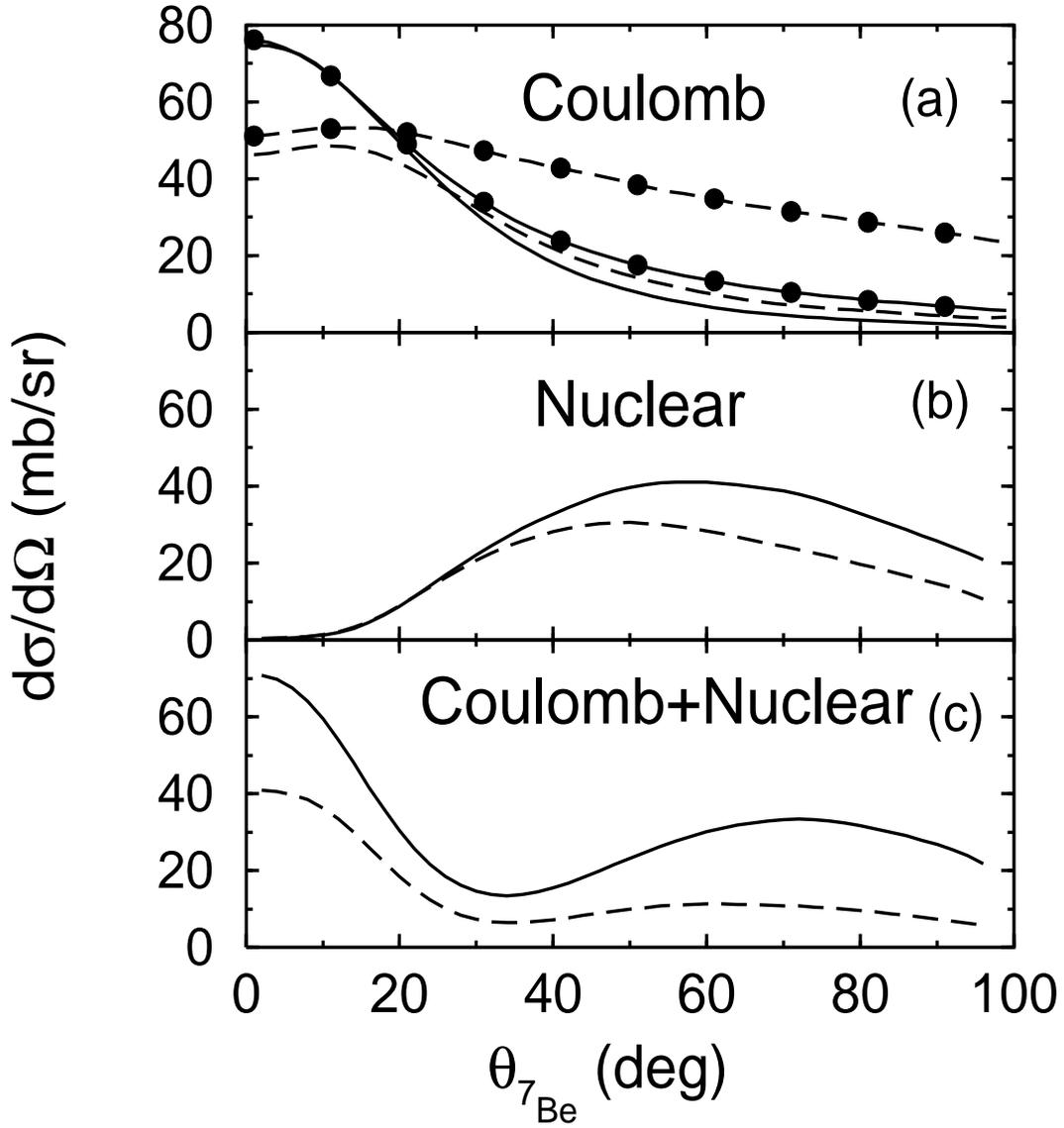,height=15.0cm,width=14.0cm}}
\end{center}
\caption{
Dipole (solid lines) and quadrupole (dashed lines) components of the
angular distributions of the $^7$Be 
fragment emitted in the breakup reaction of $^8$B on $^{58}$Ni target at
the beam energy of 25.8 MeV. (a) pure Coulomb breakup; also
shown here are the $E1$ (solid lines with solid circles) and $E2$
(dashed lines
with solid circles) cross sections obtained with point-like projectile
and target approximation (Alder-Winther theory), (b) pure
nuclear breakup and (c) Coulomb plus nuclear breakup where the
corresponding amplitudes are coherently summed.}     
\label{fig:figb}
\end{figure}

\begin{figure}
\begin{center}
\mbox{\epsfig{file=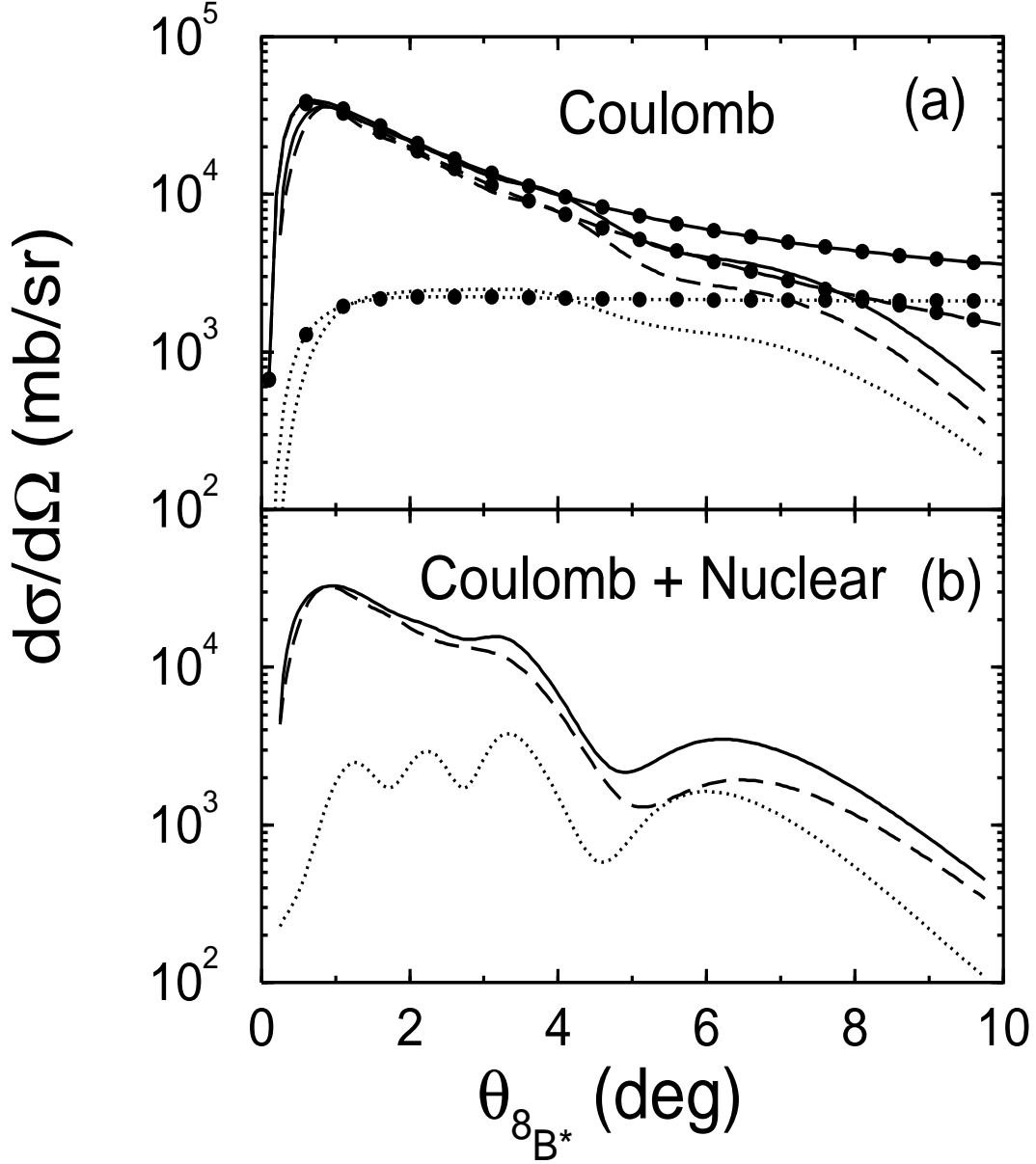,height=16.0cm,width=14.0cm}}
\end{center}
\vskip .3in
\caption{
Angular distribution for $^8$B+$^{208}$Pb $\rightarrow$ $^8$B$^*$($^7$Be+p)+
$^{208}$Pb reaction at the beam energy of 415 MeV.
(a) Results for pure Coulomb excitation, the dashed and
dotted curves represent the $E1$ and $E2$ cross sections while their sum is
depicted by the solid line. Also shown here are the results obtained with a
point-like projectile and target approximation (Alder-Winther theory), where
dashed and dotted lines with solid circles show the corresponding
$E1$ and $E2$
cross sections while the solid line with solid circles represents their
sum. (b) Coherent sum of Coulomb and Nuclear excitation calculations;
the dashed and dotted lines show the dipole and quadrupole components while
the solid line is their sum. }
\label{fig:figc}
\end{figure}

\begin{figure}
\begin{center}
\mbox{\epsfig{file=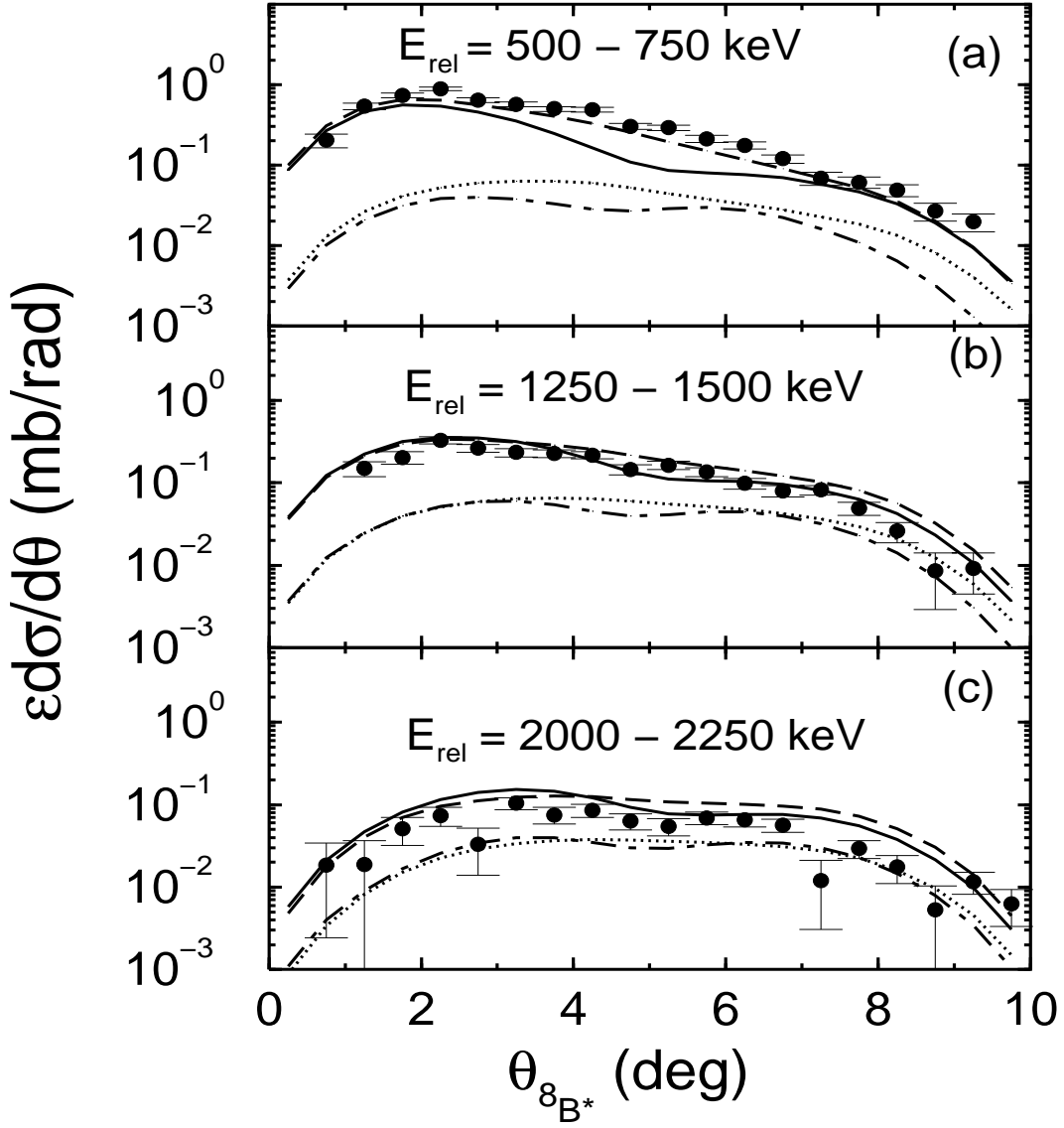,height=15.0cm,width=14.0cm}}
\end{center}
\caption{
Comparison of experimental and theoretical cross section
$\epsilon d\sigma/d\theta$ as a function of
the scattering angle $\theta_{8_{B^*}}$ 
for $^8$B+$^{208}$Pb $\rightarrow$ $^8$B$^*$($^7$Be+p)+
$^{208}$Pb reaction at the beam energy of 415 MeV.
Results for three relative energy bins of (a) 500-750 keV,
(b) 1250-1500 keV, (c) 2000-2250 keV are shown. $\epsilon$ is 
the detector efficiency. Solid lines show the calculated total Coulomb plus
nuclear dissociation cross sections while the dashed lines represents the
corresponding pure Coulomb dissociation result. Pure quadrupole  Coulomb and
Coulomb+nuclear cross sections are shown by dotted and dashed-dotted lines.
The experimental data and the detector efficiencies are taken
from~\protect\cite{moto2}.} 
\label{fig:figd}
\end{figure}
\end{document}